\documentclass[aps,preprint]{revtex4}
\usepackage{epsfig,amsmath}
\usepackage{subfigure}
\usepackage{graphicx}
\usepackage{dcolumn}
\usepackage{stmaryrd}
\usepackage{mathrsfs}
\usepackage{pifont}
\usepackage{amsthm}
\usepackage{amssymb}
\usepackage{bm}
\usepackage{latexsym}
\usepackage{hyperref}
\usepackage{color}

\begin{document}

\title{Symmetry restoration and quantumness reestablishment }

\author{Guo-Mo Zeng$^{1}$, Lian-Ao Wu$^{2,3}$, and Hai-Jun Xing$^{1}$ \footnote{Correspondence and requests for materials should be addressed to L.-A.W. (lianao\_wu@ehu.es) or G.-M. Z. (gmzeng@jlu.edu.cn)}}

\affiliation{$^{1}$College of Physics, Jilin University, Changchun 130012, China, $^{2}$Department of Theoretical Physics and History of Science, The Basque Country University (EHU/UPV), PO Box 644, 48080 Bilbao Spain, $^3$Ikerbasque, Basque Foundation for Science, 48011 Bilbao Spain}

\date{\today}

\begin{abstract}
A realistic quantum many-body system, characterized by a generic microscopic
Hamiltonian, is accessible only through approximation methods. The mean
field theories, as the simplest practices of approximation methods, commonly
serve as a powerful tool, but unfortunately often violate the symmetry of the
Hamiltonian. The conventional BCS theory, as an excellent mean field approach,
violates the particle number conservation and completely erases quantumness
characterized by concurrence and quantum discord  between different modes.
We restore the symmetry by using the projected BCS theory and the exact
numerical solution and find that the lost quantumness is synchronously reestablished.
We show that while entanglement remains unchanged with the particle numbers,
quantum discord behaves as an extensive quantity with respect to the system
size. Surprisingly, discord is hardly dependent on the interaction strengths.
The new feature of discord offers promising applications in modern quantum
technologies.

\end{abstract}


\maketitle


A quantum many-body Hamiltonian $H$ often possesses invariance under symmetry
operations, exemplified by the particle number $N$ and the angular momentum
${J^2}, {J_z}$. Proper treatments of symmetry are of great importance in
developing approximation methods for the Hamiltonian, in particular for
strongly correlated systems. Unfortunately, ubiquitous approximation methods
such as the mean field approach usually require breaking the symmetry of the
Hamiltonian. For instance, the renowned BCS theory \cite {BCS}, proposed for
superconductivity and later for nuclear system, employs simple product wave
functions of independent quasi-particles to reveal the underlying physics,
at a price of violating the particle-number conservation. The entire system
undergoes a phase transition to a symmetry-violating superfluid phase. In
condensed matter physics, the particle-number fluctuation of a bulk superconductor
is weak and the symmetry breaking does not play a significant role. On the
contrary, the violation manifests itself in finite-size ensembles such as
superconductive grains \cite {Grain} and nuclear systems. For instance, the
nuclear properties obtained in the BCS treatment are the average of the target
nucleus and its adjacent nuclei. The BCS theory successfully captures the
dominant quantum correlation, i.e. the pairing effect between single particle
(electron or nucleon) states and their time-reversals, but thoroughly washes
out other correlations such as the correlations between different modes. These
correlations might not be important for bulk superconductors but apparently
exist in finite-size systems, and can be restored by the symmetry-restoration
methods that go beyond the mean field approaches.

In quantum information theory, quantum and classical correlations can be
distinguished specifically. Quantum correlations characterize quantumness,
measured by concurrence and quantum discord, of a correlated system. It should
be of great interest how quantumness varies with the process from violation to
restoration of symmetry. Here we study a two-level BCS model. The conventional
BCS theory of the model violates the particle-number conservation. We shall
restore the conservation by the projected BCS (PBCS) method, which projects
out states with fixed particle numbers from the BCS wave function. It is
noticeable that the lost quantumness is found when the symmetry is restored.
Quantumness and symmetry are destroyed and reestablished simultaneously. We
study the finite particle number effects on both concurrence and quantum discord
in the case of large degeneracy. To our surprise, while entanglement drops greatly
with the particle number from the maximum at two end points and then remains at
half value, quantum discord behaves as an extensive quantity for a target system
with a fixed degeneracy. This implies that the paring interaction is effectively
transparent to discord. Physically, discord displays quantumness where a pair of
qubits moves independently as if other qubits do not exist, whereas the existence
of other qubits severely affects the entanglement. According to their definitions,
quantum discord is the quantumness associated with quantum measurement, e.g. von
Neumann measurement, on a system, while entanglement is from the density matrix
itself of the system. The difference is crystal clear. Quantum discord seems
to be a more general and fundamental nonclassical correlation than entanglement
as pointed out in recent literatures \cite {Ollivier, Dat08, Maz09, Fer10}. As
proved in Ref. \cite {Fer10}, the set of states with zero-discord has volume zero
in the whole Hilbert space, very different from the set of non-entangled states
which have finite volumes. Since almost all quantum states may be useful resources
for discord-dependent tasks due to their positive discord, our result that discord
acts as an extensive quantity, can be important in assessing availability of a
quantum state in various quantum information tasks.

The BCS theory \cite{BCS} was proposed to determine the fully paired ground
state of superconductors and later was translated into the theories of nuclear
structure \cite{Bayman} since the pairing correlation was found in low-lying
states of even-even nuclei. In the past decade some authors studied the entanglement
of superconductors in the frame of BCS model considering that it provides a solution
to a quantum many-body problem with an explicit wave function. Ref. \cite{Martin}
used concurrence to quantify the entanglement of the BCS state of
superconducting compounds. Ref. \cite{Dusuel} analyzed the finite-size properties
of the two-level BCS model and discussed the entanglement properties of the ground
state via concurrence. Without loss of generality, we now focus on the BCS application
in nuclear systems, however, the conclusions obtained can be applied to any finite
systems directly.

The ground-state BCS wave function of an even-even nucleus reads \cite{Ring}
\begin{equation}\label{eq:1}
 \mid{\textrm{BCS}}\rangle = \prod_{k>0}(u_k + v_k a_k^{\dagger}a_{\bar{k}}^{\dagger})|0 \rangle,
\end{equation}
which violates particle-number conservation. Here $a_k^{\dagger}$ and
$a_{\bar k}^{\dagger}$ are fermionic creation operators of single particle
state $k$ and its time-reversal ${\bar k}$, respectively. $u_k$ and $v_k$
are real variational parameters satisfying the normalization condition
$u_k^2 + v_k^2 = 1$. These parameters are determined by variation of the
energy, with the restrictions that the expectation value of particle number
operator $\hat N$ is equal to the practical particle number $N$, i.e.,
$\langle{\rm{BCS}}| \hat N |{\rm{BCS}}\rangle=N$.

The pure pairing interaction is widely used in nuclear BCS theory, for it
provides a simple and powerful description of paring correlations in nuclei.
A generic Hamiltonian with the pairing interaction reads
\begin{equation}\label{eq:2}
 H = \sum\limits_{k>0}   {\epsilon_k\left(a_k^{\dagger}a_k + a_{\bar k}^{\dagger}a_{\bar k} \right)}
   - \sum\limits_{k,k'>0}{G_{kk'kk'}        a_k^{\dagger}a_{\bar k}^{\dagger}a_{\bar k'}a_{k'}},
\end{equation}
where ${\epsilon _k}$'s are single-particle energies and $G_{kk'kk'}$ are the
strengths of paring force dependent on the energy levels. The BCS theory is
convenient and widely used. However it violates the particle number conservation
and the physics obtained from BCS treatment is an average of the target nucleus
and its adjacent nuclei. This symmetry breaking can be restored by the PBCS theory
\cite{DMP}, as introduced in {\textbf {Method}}.

We now construct three bilinear fermionic operators,
$S_+^{(k)} = a_k^{\dagger}a_{\bar k}^{\dagger}$, $S_-^{(k)} = a_{\bar k}a_k$ and
$S_0^{(k)} = \textstyle{\frac{1}{2}}\left({a_k^{\dagger}{a_k}+a_{\bar k}^{\dagger}{a_{\bar k}}-1} \right)$,
which play the same role as spin operators (half of Pauli operators) and
generate an $su(2)$ algebra. The operator $S_0$ has eigenvalues $\pm \frac{1}{2}$
depending on whether the pair state $(k,\bar k )$ is occupied or not. We denote
the empty and occupied with $|0\rangle$ and $|1\rangle$, two states supporting
a qubit. The equivalence between fermonic pairs and qubits is discussed in Refs.
\cite{WuBCS, WuPara,WuPara1}. Specifically, a fermonic pair in $(k,\bar k)$ corresponds
to the $k$th qubit.

\bigskip
\noindent\textbf{\large {Results}}\\
\noindent Consider a simplified pairing Hamiltonian \eqref{eq:2} (similar
to the nuclear model in Ref. \cite{WuC,WuC1}): two degenerate single particle
energy levels, set as $\epsilon_1 = 0$ and $\epsilon_2 = 1$, with the
degeneracies $\Omega_1$ and $\Omega_2$, respectively. The strengths of
paring force are $G_{k\bar{k}k'\bar{k'}}=4G_{ii}$ $(i=1,2)$ for both $k$
and $k'$ in the same level $i$ and $G_{k\bar{k}k'\bar{k'}}=4 G_{12}=4G_{21}$
for $k$ and $k'$ in different levels. We focus on the low-lying states
where all particles are coupled in pairs. The BCS and PBCS wave functions
are given by Eqs. (\ref{eq:1}) and (\ref{eq:5}) in {\textbf{Methods}.

Eq. \eqref{eq:1} can be rewritten such that the BCS wave function becomes
a direct product state of qubit states of different modes. Both concurrence
and quantum discord are therefore vanishing. It shows that while violating
particle number conservation, BCS wave function annihilates quantumness of
different modes.

Now we calculate the reduced density matrix of two qubits $A$ and $B$ under
the PBCS wave function \cite{Ring}. The Hilbert space of the two qubits
is spanned by the computational base, $\left\{|00\rangle, |01\rangle, |10\rangle, |11\rangle \right\}$,
where in each basis, the first (second) digit indicates the state of qubit $A(B)$.
There are three different types of reduced density matrices, determined by
the way qubits $A$ and $B$ occupy the two energy levels, and all of them
are a particular case of two-qubit $X$ states \cite{Wootters,Sar09,Luo,Ali,FeiSM,FanH1,FanH2} with
$\rho_{14}=\rho_{41}=0$, which in general reads
\begin{equation}\label{eq:3}
\rho = {\left(
\begin{array}{cccc}
\rho_{11} &     0     &     0     &     0    \\
    0     & \rho_{22} & \rho_{23} &     0    \\
    0     & \rho_{32} & \rho_{33} &     0    \\
    0     &     0     &     0     & \rho_{44}\\
\end{array}
\right)},
\end{equation}

We distinguish the three types by superscripts (1), (2) and (3). Type (1)
denotes the case that both qubits $A$ and $B$ are in the lower level; (2)
A and B in the lower and the higher levels, respectively; (3) both $A$
and $B$ in the higher level. The non-zero matrix elements of $\rho^{(1)}$
are:
\begin{align}\label{eq:13}
 \nonumber \rho_{11}^{(1)} & = C\sum_{i=\max\{0,p-\Omega_2\}}^{\min\{p,\Omega_1-2\}}{\binom{\Omega_1-2}{i}
   \left({\frac{v_1^2}{u_1^2}}\right)^i\binom{\Omega_2}{p-i}\left({\frac{v_2^2}{u_2^2}}\right)^{p-i}}, \\
 \nonumber \rho_{22}^{(1)}&=\rho_{33}^{(1)}= \rho_{23}^{(1)}=\rho_{32}^{(1)} \\
                   & = C\sum_{i=\max\{1,p-\Omega_2\}}^{\min\{p,\Omega_1-1\}}{\binom{\Omega_1-2}{i-1}
 \nonumber \left({\frac{v_1^2}{u_1^2}}\right)^i\binom{\Omega_2}{p-i}\left({\frac{v_2^2}{u_2^2}}\right)^{p-i}}, \\
           \rho_{44}^{(1)} & = C\sum_{i=\max\{2,p-\Omega_2\}}^{\min\{p,\Omega_1\}}{\binom{\Omega_1-2}{i-2}
   \left({\frac{v_1^2}{u_1^2}}\right)^i\binom{\Omega_2}{p-i}\left({\frac{v_2^2}{u_2^2}}\right)^{p-i}},
\end{align}
where $ \left({\begin{array}{*{20}{c}} n \\ k \\ \end{array}}\right)$ stands for
the number of combination, the common factor $C=u_1^{2\Omega_1}u_2^{2\Omega_2}$.
For the second type, the matrix elements are
\begin{align}\label{eq:14}
 \nonumber \rho_{11}^{(2)} & = C\sum_{i=\max\{0,p-\Omega_2+1\}}^{\min\{p,\Omega_1-1\}}{\binom{\Omega_1-1}{i}
   \left({\frac{v_1^2}{u_1^2}}\right)^i\binom{\Omega_2-1}{p-i}  \left({\frac{v_2^2}{u_2^2}}\right)^{p-i}}, \\
 \nonumber \rho_{22}^{(2)} & = C\sum_{i=\max\{0,p-\Omega_2\}}^{\min\{p-1,\Omega_1-1\}}{\binom{\Omega_1-1}{i}
   \left({\frac{v_1^2}{u_1^2}}\right)^i\binom{\Omega_2-1}{p-i-1}\left({\frac{v_2^2}{u_2^2}}\right)^{p-i}}, \\
 \nonumber \rho_{33}^{(2)} & = C\sum_{i=\max\{0,p-\Omega_1\}}^{\min\{p-1,\Omega_2-1\}}{\binom{\Omega_1-1}{p-i-1}
   \left({\frac{v_1^2}{u_1^2}}\right)^{p-i}\binom{\Omega_2-1}{i}\left({\frac{v_2^2}{u_2^2}}\right)^{i}},   \\
 \nonumber \rho_{23}^{(2)} & =\rho_{32}^{(2)}=\sqrt{\rho_{22}^{(2)} \rho_{33}^{(2)}}, \\
           \rho_{44}^{(2)} & =C\sum_{i=\max\{1,p-\Omega_2\}}^{\min\{p-1,\Omega_1\}}{\binom{\Omega_1-1}{i-1}
   \left({\frac{v_1^2}{u_1^2}}\right)^i\binom{\Omega_2-1}{p-i-1}\left({\frac{v_2^2}{u_2^2}}\right)^{p-i}}.
\end{align}
The reduced density matrix of third type is similar to first one. When
$\Omega_1=\Omega_2=\Omega$, the two types are connected by a simple relationship,
$\rho^{(1)}(p) = \rho^{(3)}(2\Omega-p)$, due to the particle-hole symmetry with
$2\Omega-p$ being the hole number.

For the X-type density matrix, one can derive the expression of
concurrence as \cite{Wootters}
\begin{equation}\label{eq:15}
\mathbb{C}(\rho)=\max \left\{0,2\sqrt{\rho_{22}\rho_{33}}-2\sqrt{\rho_{11}\rho_{44}}\right\}.
\end{equation}
The total correlation of the system can be written as \cite{Sar09}
\begin{align}\label{eq:16}
 \nonumber {\cal I}(\rho)= & \rho_{11}\log_2\rho_{11}+\rho_{44}\log_2\rho_{44}+\left(\rho_{22}+\rho_{33}\right)\log_2\left(\rho_{22}+\rho_{33}\right) \\
 & -\left(\rho_{11}+\rho_{22}\right)\log_2\left(\rho_{11}+\rho_{22}\right)-\left(\rho_{33}+\rho_{44}\right)\log_2\left(\rho_{33}+\rho_{44}\right) \\
 \nonumber
 & -\left(\rho_{11}+\rho_{33}\right)\log_2\left(\rho_{11}+\rho_{33}\right)-\left(\rho_{22}+\rho_{44}\right)\log_2\left(\rho_{22}+\rho_{44}\right).
\end{align}
The classical correlation is defined, based on the von Neumann measurements
$\left\{B_k\right\} (k=0,1)$ on subsystem $B$, as
\begin{align}\label{eq:17}
 {\cal C}(\rho) =& -\left(\rho_{11}+\rho_{22}\right)\log_2\left(\rho_{11}+\rho_{22}\right) \\
 \nonumber       & -\left(\rho_{33}+\rho_{44}\right)\log_2\left(\rho_{33}+\rho_{44}\right)-\min \left\{S_1,S_2\right\},
\end{align}
where we follow the discussions of Ref. \cite{Ali} and here
\addtocounter{equation}{1}
\begin{align}\label{eq:18}
 \nonumber S_1 = & -\rho_{11}\log_2\frac{\rho_{11}}{\rho_{11}+\rho_{33}}
                   -\rho_{33}\log_2\frac{\rho_{33}}{\rho_{11}+\rho_{33}} \\
                 & -\rho_{22}\log_2\frac{\rho_{22}}{\rho_{22}+\rho_{44}}
                   -\rho_{44}\log_2\frac{\rho_{44}}{\rho_{22}+\rho_{44}}
\end{align}
and
\begin{align}\label{eq:19}
 S_2=& -\frac{1-\theta}{2}\log_2\frac{1-\theta}{2}-\frac{1+\theta}{2}\log_2\frac{1+\theta}{2}
\end{align}
with
\begin{equation}\label{eq:20}
 \theta=\sqrt{\left(\rho_{11}+\rho_{22}-\rho_{33}-\rho_{44}\right)^2+4\rho_{22}\rho_{33}}
\end{equation}

\noindent{\textbf{Two-level case.}} We now consider a pair-correlated system with two degenerate
levels. We first compute concurrence and quantum discord based on PBCS for
a simple case that $\Omega_1=\Omega_2=\Omega, G_{11}=G_{12}=G_{21}=G_{22}=G$.
The concurrence and quantum discord of three types versus pairing number $p$
are shown in Fig. \ref{fig1} (a) and (b), respectively. At first glance, we
notice that both curves of concurrence and quantum discord exhibit symmetry
about $p=\Omega$, perfectly for the second type, and approximately for the first
and third types, as well as the particle-hole symmetry between first case
and third case as mentioned above. Thus, in the following we only need to
concentrate our discussion on the region $0 \le p \le \Omega$ and one of
the three types, e.g., type 2. The more meaningful and essential result
displayed in Fig. \ref{fig1} is that concurrence and discord exhibit totally
different behaviors. From Fig. \ref{fig1} (a) we can see that concurrence
starts with zero at $p=0$ corresponding to the BCS ground state, jumps to
the maximal value at $p=1$, then decreases by near half and keeps almost
unchanged with particle number. Clearly, it shows that symmetry restoration
results in the reestablishment of entanglement.

On the other hand, Fig. \ref{fig1} (b) shows that quantum discord also starts
with zero, indicating that the BCS state is not only unentangled but also
classic-only correlated. The different behavior of discord occurs after $p=1$,
it does not drop like concurrence but increases with pair numbers until its
highest value as the total maximal occupation number of the two levels are
half-filled. This fact implies that concurrence and discord are different
aspects of quantumness, owing to their distinct nature. There have been some
reports about physical phenomena where entanglement and discord behave differently.
A typical example is that discord can indicate the Kosterlitz-Thouless phase
transition in the XXZ model, whereas concurrence cannot \cite{Sar09, MPM10, MCSS12}.

In Fig.\ref{fig2} we plot together the curves of concurrence and discord, as
well as the total correlation $\mathcal{I}(\rho)$ and classical correlation
$\mathcal{C}(\rho)$ (refer to Ref. \cite{FeiSM} for detailed expressions) vs
particle number for the second type with $\Omega_1=\Omega_2=20$. Apart from
the remarkable difference between concurrence and quantum discord, we can
also see that $\cal{D} \ge \cal{C}$ for the whole range of particle number,
oppositing to the early speculation that $\cal{C} \ge \cal{D}$ for any
quantum state \cite{GPW05, HV0103, Horo05}.

The Hamiltonian (\ref{eq:2}) can be exactly diagonalized. The numerical
calculations show that the differences between PBCS and the exact solution
will diminish with degeneracy and tend to vanish as the degeneracy is large
enough. For a small degeneracy, e.g., $\Omega_1=\Omega_2=6$, the fully restored
quantum correlations measured by concurrence and discord in the exact solution
are slightly larger than those in PBCS. It shows that PBCS is a perfect
approximation to the exact solution in the sense of restoring the quantum
correlations. However, superior to PBCS method, the exact diagonalization
can even treat more complicated cases. In particular, we set
$G_{11}=0.7,G_{12}=0.6, G_{22}=0.5$ for the fixed $\Omega_1 = \Omega_2 =6$,
and $\Omega_1 = 7, \Omega_2 =5$ for the fixed $G_{11}=G_{12}=G_{22}=0.6$,
respectively. Fig. \ref{fig3} shows concurrence and quantum discord, and
compare with {\em a typical case} that $\Omega_1 = \Omega_2 =6$ and $G_{11}=G_{12}=G_{22}=0.6$.
The deviations from {\em the typical case} are slight.

Comparing Fig. 2 and Fig. 3, we also notice that for larger degeneracy, quantum
discord increases approximately linearly with particle (hole) pair, when
$p, q \ll \Omega$, indicating that it acts as an {\em extensive quantity} like
internal energy. We calculate the cases when $\Omega_1\neq\Omega_2$, and
$G_{11} \neq G_{22} = G_{33} \neq G_{44}$, and confirm that these interesting
properties remain the same. From Fig. 2, we can see that the quantum mutual
information, i.e., the total correlation displays a much better linearity than
discord. For $p \le 8$ the variation of total correlation versus
$p$ is very close to a straight line with slope $\sim 0.42$. For $p \le 4$
the quantum discord also coincides quite well with the straight line. An
extensive quantity normally is the sum of the properties of separate noninteracting
subsystems that compose the entire system. This implies astonishingly that
the pairing interaction is effectively {\em transparent} to discord. For
instance, discord as a function of the pairing strength $G$: $\mathcal{D}(G)\rightarrow$ 1.006(0.4), 1.003(0.5), 0.998(0.7),
0.997(0.8), 0.996(0.9) and 0.995(1.0) in unit of $\mathcal{D}(G=0.6)=0.03168$
when $\Omega=120$ and $p=4$. The values of $\mathcal{D}(G)$ merely change
with $G$, in particular when $G$ is strong. Physically, discord displays
quantumness where a pair of qubits {\em moves} independently as if other
qubits do not exist, whereas the existence of other qubits severely affects
the values of entanglement. Based on their definitions, discord reflects the
quantumness from quantum measurement on the interested system, meanwhile
entanglement is from the density matrix directly of the system. In comparison
with entanglement, discord quantifies more general and more fundamental quantum
correlations \cite {Ollivier, Dat08, Maz09, Fer10}. In particular, Ref. \cite{Fer10}
shows that the set of states with vanishing discord, i.e., the set of classical
states, has volume zero in the whole Hilbert space. This is very different from
the set of separable (non-entangled) states, which exhibits finite volume. This
result provides another perspective for understanding the differences between
entanglement and discord. Considering that the existence of discord determines
the nontrivial properties of quantum states, it was shown that discord may be
the resource responsible for the quantum speedup \cite {Dat08} in computational
models. This implies that almost all quantum states are useful resources due
to their positive discord. The feature, discord as an extensive quantity, is
expected to have new applications in quantum information practices.

\noindent{\textbf{One-level case.} In order to understand the above results
deeply, we consider a one-level system with degeneracy $\Omega$.
This is the limit of the two-level cases when level difference vanishes.
The density matrix now reads
\begin{equation}\label{eq:21}
\rho = \frac{1}{\Omega(\Omega-1)}
\left({\begin{array}{*{20}{c}}
 q(q-1) & 0  & 0  & 0 \\
      0 & pq & pq & 0 \\
      0 & pq & pq & 0 \\
      0 & 0  & 0  & p(p-1) \\
\end{array}} \right)
\end{equation}
where $q=\Omega-p$ is the hole-pair number in the single level. The strength of
paring force $G$ does not appear in the density matrix, which may be the reason
why in the two-level case the entanglement and correlations are hardly affected
by the strength of paring force. It turns out that it is paring correlation itself,
rather than paring interaction, that determines these novel properties of a
pair-correlated many-body system. Here we emphasize that $\rho$ is symmetric
under the interchange of $p$ and $q$, or  interexchange of states $\left|00\right\rangle$
and $\left|11\right\rangle$, so are the entanglement, classical correlations
and quantum discord. In one-level case, all correlations have explicit analytical
expressions, the concurrence is
\begin{equation}\label{eq:22}
 {\mathbb C}(\rho)= \frac{2}{\Omega(\Omega-1)}\left[pq-\sqrt{p(p-1)q(q-1)}\right],
\end{equation}
and the total correlation reads
\begin{align}\label{eq:23}
 {\cal I}(\rho) = \frac{p(p-1)}{\Omega(\Omega-1)}\log_2\frac{p(p-1)}{\Omega(\Omega-1)}
 \nonumber       +\frac{q(q-1)}{\Omega(\Omega-1)}\log_2\frac{q(q-1)}{\Omega(\Omega-1)} \\
                 +\frac{2pq}   {\Omega(\Omega-1)}\log_2\frac{2pq}   {\Omega(\Omega-1)}
                 -\frac{2p}{\Omega}\log_2\frac{p}{\Omega}
                 -\frac{2q}{\Omega}\log_2\frac{q}{\Omega},
\end{align}
As for classical correlation, we compare
\begin{align}\label{eq:24}
 \nonumber S_1 = & -\frac{pq}    {\Omega(\Omega-1)}\log_2\frac{q}  {\Omega-1}
                   -\frac{p(p-1)}{\Omega(\Omega-1)}\log_2\frac{p-1}{\Omega-1} \\
                 & -\frac{pq}    {\Omega(\Omega-1)}\log_2\frac{p}  {\Omega-1}
                   -\frac{q(q-1)}{\Omega(\Omega-1)}\log_2\frac{q-1}{\Omega-1}
\end{align}
and
\begin{align}\label{eq:25}
 S_2 = -\frac{1-\theta}{2}\log_2\frac{1-\theta}{2}
       -\frac{1+\theta}{2}\log_2\frac{1+\theta}{2}
\end{align}
with
\begin{equation}\label{eq:26}
 \theta = \frac{1}{\Omega(\Omega-1)}\sqrt{(p-q)^2(\Omega-1)^2+4p^2q^2}
\end{equation}
We find that $\min \left\{S_1, S_2\right\}=S_2$, thus the classical correlation
is given by
\begin{align}
 \nonumber {\cal C}(\rho) = &  \frac{1-\theta}{2}\log_2\frac{1-\theta}{2}
                              +\frac{1+\theta}{2}\log_2\frac{1+\theta}{2} \\
                            & -\frac{p}{\Omega}\log_2\frac{p}{\Omega}
                              -\frac{q}{\Omega}\log_2\frac{q}{\Omega}
\end{align}
and thereby the quantum discord is
\begin{align}\label{eq:27}
 {\cal D}(\rho)= &  \frac{p(p-1)}{\Omega(\Omega-1)}\log_2\frac{p(p-1)}{\Omega(\Omega-1)}
 \nonumber         +\frac{q(q-1)}{\Omega(\Omega-1)}\log_2\frac{q(q-1)}{\Omega(\Omega-1)} \\
                 & +\frac{2pq}{\Omega(\Omega-1)}\log_2\frac{2pq}{\Omega(\Omega-1)}
 \nonumber         -\frac{p}{\Omega}\log_2\frac{p}{\Omega}
                   -\frac{q}{\Omega}\log_2\frac{q}{\Omega} \\
                 & -\frac{1-\theta}{2}\log_2\frac{1-\theta}{2}
                   -\frac{1+\theta}{2}\log_2\frac{1+\theta}{2}.
\end{align}
These expressions are exactly symmetric under the interchange of $p$ and $q$,
therefore all of them are even functions of $p-q$ and their extrema lie at
$p=q=\Omega/2$ and at the ends. In what follows we will focus our discussions
on particle pair $p$ due to the symmetry.  All results are the same once
replacing $p$ with $q$.  Concurrence $\mathbb C$ is a monotone decreasing
function of $p$ for $p<\Omega/2$ and reaches its minimum at $p=\Omega/2$.
However, for a fixed degeneracy $\Omega$,
the values of concurrence hardly vary with $p$, as shown in the above equations.
Concurrence has the limit, as $\Omega \gg 1,{\rm{ }}p \gg 1,{\rm{ }}p \ll \Omega$,
\begin{equation}\label{eq:29}
  {\mathbb C}(\rho) \simeq \frac{1}{\Omega-1},
\end{equation}
which is the same as its minimum at $p=\frac{\Omega}{2}$. This indicates that
concurrence behaves as an intensive quantity. On the other hand, concurrence
vanishes at large $\Omega$ because of the factor $\frac{2}{\Omega(\Omega-1)}$
in Eq. (13). This result may be used to choose materials as the resource of
quantumness according to their degeneracy and the number of particle (or hole)
pairs.

On contrary to concurrence, the total correlation, classical correlation and
quantum discord behave completely differently. They are monotonically increasing
functions of $p$ in the intervals $p \in \left({0,{\rm{ }}\Omega /2}\right)$
with $p=\Omega/2$ as the maximum. The maximal values are
\begin{align}\label{eq:30}
 &{\cal I}_{\rm {max}}=\log_2\frac{\Omega-2}{\Omega-1}+\frac{\Omega}{{2\left(\Omega-1\right)}}\log_2\frac{2\Omega}{\Omega-2} \\
 &{\cal C}_{\rm {max}} =\frac{1}{4}\left(\frac{ \Omega-2}{\Omega-1}\log_2\frac{\Omega-2}{ \Omega-1}+\frac{3\Omega-2}{\Omega-1}\log_2\frac{3\Omega-2}{\Omega-1}\right)-1 \\
 &{\cal D}_{\rm {max}} =\frac{1}{4}\left(\frac{3\Omega-2}{\Omega-1}\log_2\frac{\Omega-2}{3\Omega-2}+\frac{2\Omega  }{\Omega-1}\log_2\frac{2\Omega  }{\Omega-2}\right)+1
\end{align}
When $\Omega \gg 1$, their limits are given by
\begin{align}\label{eq:31}
 &{\cal I}_{\rm {max}} \to \frac{1}{2}, \\
 &{\cal C}_{\rm {max}} \to \frac{3}{4}\log_2 3-1 \simeq 0.189, \\
 &{\cal D}_{\rm {max}} \to \frac{3}{2} - \frac{3}{4}{\log _2}3 \simeq 0.311,
\end{align}
It means that no matter how large the degeneracy is, the above correlations do not vanish but remain a constant.

Under the large degeneracy limits with small numbers of pairs, the total correlation, classical correlation and
quantum discord are
\begin{align}\label{eq:32}
 & {\left. \cal I\left(\rho\right) \right|_{\Omega \gg 1,{\rm{ }}p \ll \Omega}} \simeq  \frac{2p}{\Omega}, \\
 & {\left. \cal C\left(\rho\right) \right|_{\Omega \gg 1,{\rm{ }}p \ll \Omega}} \simeq -\frac{ p}{\Omega}\log_2\frac{p}{\Omega}, \\
 & {\left. \cal D\left(\rho\right) \right|_{\Omega \gg 1,{\rm{ }}p \ll \Omega}} \simeq  \frac{2p}{\Omega}+\frac{p}{\Omega}\log_2\frac{p}{\Omega}.
\end{align}
It is thus clear that the total correlation increases linearly with particle
pair, while the quantum discord increases almost linearly, as observed in the
two-level case where the slope is $\sim 0.42$ when $\Omega =40$.
We therefore further confirm, by the analytical one-level case that the conclusions
of discord being an extensive quantity in the two-level case remains valid and
should imply profound universality for any many-body system consisting of pairwise
correlated particles. These theoretical results are obviously testable experimentally.
For example strong correlated many-body nuclear systems may allow to choose a series
of even-even isotones, which usually have the same single particle levels for valence
neutrons, to examine the dependance of quantum correlations on the number of paired
neutrons. We can also select a group of nuclei which have the same number of valence
protons (or neutrons) but different valence shells to check our results for quantum
correlations.

\bigskip
\noindent\textbf{\large {Discussion}}

\noindent We have studied symmetry restoration and quantumness reestablishment
of the BCS theory, as well as the relationship between paring correlation and
quantumness in pair-correlated many-body systems. Restored entanglement may
not matter for bulk superconductors because it keeps on a low level with
particle-pair numbers. Quantum discord, on the other hand, is an extensive
quantity and grows linearly with pair numbers. From the perspective of paring
interaction, a pair of qubits moves independently, while surrounding qubits
strongly affects entanglement of the target qubit pair. The underlying origin
of these effects should be from that discord is a measurement-dependent quantity,
while entanglement is only determined by the density matrix of the target system.
Besides, because of the volume-zero of the set of classic states, these states
can be hardly found in Hilbert space. Considering that in many physical systems,
quantum discord behaves qualitatively similar to entanglement, it is significant
to find physical entities where discord substantially differs from entanglement.
An example is that discord signals KT phase transition while entanglement
cannot. Our results display a new aspect of the differences, with respect
to particle number effect, which may play a unique role in exploring strongly
correlated systems like high temperature superconductors since the quantity is hardly
dependent on interactions in strong interaction regime \cite {WuO5}. The new
feature may hint profound physics behind quantum discord and may offer promising
applications for, e.g., certain quantum computations.

It is worth mentioning that quantum discord can also be discussed in a more
general setup of positive operator valued measurements (POVM)\cite{Ali}
\cite{ChenQ}. We can expect that, by considering POVM instead of von Neumann
measurement, the main results will remain essentially the same. For X-states
as considered here, extremization via POVM can be reduced to orthogonal projectors
for a number of states. \cite{ChenQ}

\bigskip
\noindent\textbf{\large {Methods}}\\
\noindent\textbf{BCS and PBCS.} By introducing the Lagrange multiplier $\lambda$,
called the chemical potential,  we determine the BCS parameters by variation of
the BCS expectation value of $H'=H-{\lambda}\hat N$
\begin{align}\label{eq:3}
 \langle{\rm{BCS}}|H'|{\rm{BCS}}\rangle =& \sum\limits_{k>0}{\left(\epsilon_k - \lambda \right)v_k^2}
 \nonumber    + \sum\limits_{k,k'>0}{G_{kk'kk'}v_k^2 v_{k'}^2} \\
            & + \sum\limits_{k,k'>0}{G_{k\bar kk'\bar k'}u_k v_k u_{k'} v_{k'}},
\end{align}
and yields two quadratic equations for $u_k^2$ and $v_k^2$,
\begin{equation}\label{eq:4}
 u_k^2 = \frac{1}{2}\left(1-\frac{\tilde \epsilon_k}{\sqrt{\tilde \epsilon_k^2 + \Delta_k^2}} \right),{\rm{}}
 v_k^2 = \frac{1}{2}\left(1+\frac{\tilde \epsilon_k}{\sqrt{\tilde \epsilon_k^2 + \Delta_k^2}} \right),
\end{equation}
where ${\tilde \epsilon_k} = {\epsilon_k} + \sum\limits_{k'>0}(G_{kk'kk'}
+ G_{\bar kk'\bar kk'})v_{k'}^2-\lambda$ and the gap parameters
$\Delta_k = -\sum\limits_{k'>0}G_{k\bar kk'\bar k'}u_{k'}v_{k'}$. The BCS
method provides the convenience that we can treat a nucleus as a system of
quasi-particles independently moving in a mean field. However it violates
the particle number conservation.

This symmetry breaking can be restored by projection techniques, of which the
method of residues \cite{DMP} employs a projector
${\hat P^A} = \frac{1}{2\pi i}\oint {\frac{z^{\hat N}}{z^{A+1}}} dz$
to act on the BCS wave function Eq.(1)
\begin{equation}\label{eq:5}
 \left| \Psi^N \right\rangle = \hat P^{N=2p} \left| \Phi \right\rangle
  = \frac{1}{2\pi i}\oint{ \frac{d\zeta}{\zeta^{p+1}}\prod\limits_k{ \left(u_k + v_k\zeta{a_k}^{\dagger}{a_{\bar k}}^{\dagger} \right)}
    \left|0\right\rangle },
\end{equation}
where $\zeta  = {z^2}$ and $p = N/2$ is the number of particle pairs. The
integrand in the above equation is a Laurent series in $\zeta$. Making use
of the fermionic commutation relations for the operators ${a_k},{a_k}^{\dagger}$,
the matrix elements of an observable can be expressed by the residues
\begin{equation}\label{eq:6}
 R_v^m\left(k_1, \cdots, k_m \right) = \frac{1}{2\pi i}\oint {\frac{dz}{z^{(p-v)+1}}
 \prod\limits_{\scriptstyle k \ne {k_1}, \cdots ,{k_m} \hfill
 \frac \scriptstyle {\rm{      }} > 0 \hfill} {\left(u_k^2+zv_k^2\right)} }.
\end{equation}
For example, the expectation value of energy is written as,
\begin{equation}\label{eq:7}
\begin{array}{l}
 E_{\rm{proj}}^A = \frac{\left\langle \Psi^A \right| H \left| \Psi^A \right\rangle}
                        {\left\langle \Psi^A           \left| \Psi^A \right. \right\rangle }
  = \left(R_0^0\right)^{-1} \left\{ {\sum\limits_{k > 0}{\varepsilon_k} v_k^2 R_1^1(k)} \right. \\
    \left. +\sum\limits_{k,k'>0} {\left[G_{kk'kk'}v_k^2 v_{k'}^2 R_2^2 \left(k,k'\right)
    +G_{k\bar kk'\bar k'}u_k v_k u_{k'} v_{k'} R_1^2 \left(k,k'\right) \right]} \right\}. \\
\end{array}
\end{equation}
This projection method is termed as the PBCS theory.

\noindent\textbf{Concurrence and discord.} Consider a mixed state $\rho$ of
two qubits $A$ and $B$. By introducing operator
${\tilde {\rho}} = (\sigma_y \varotimes \sigma_y ) {\rho^{*}}(\sigma_y \varotimes \sigma_y)$
\cite{Wootters}, where $\sigma_y$ is the $y$ component of Pauli matrix,
one defines the concurrence of $\rho$,
\begin{equation}\label{eq:8}
 \mathbb{C}(\rho)=\max \left\{0,\sqrt{\lambda_1}-\sqrt{\lambda_2}-\sqrt{\lambda_3}-\sqrt{\lambda_4}\right\},
\end{equation}
where $\lambda_{i}$ $({i} = 1, 2, 3, 4 )$ are the eigenvalues of $\rho\tilde{\rho}$
in descending order. When $\mathbb{C}(\rho)> 0$, qubits $A$ and $B$ are entangled.

Quantum discord, as another kind of correlation, may exist even without entanglement
\cite{Ollivier}. The quantum correlation features itself with many aspects, e.g.,
in characterizing quantum phase transitions \cite{Sar09}. For the state $\rho$,
quantum discord reads:
\begin{equation}\label{eq:9}
 {\cal D}(\rho) = \cal{I}(\rho)-\cal{C}(\rho),
\end{equation}
where $\cal{I}(\rho)$ is the quantum analogue of classical mutual information,
defined as \begin{equation}\label{eq:10}
 {\cal I}(\rho) = S(\rho^A)+S(\rho^B)-S(\rho),
\end{equation}
with $\rho^{A(B)} =\textrm{Tr}_{B(A)}(\rho)$ denoting the reduced density matrix
of the partition $A(B)$, $S(\rho)$ the corresponding von Neumann entropy.
$\cal{I}(\rho)$ is interpreted as a measure of total correlations in the composite
system $A+B$, while $\cal{C}(\rho)$ is a measure of classical correlations, defined
as
\begin{align}\label{eq:11}
 {\cal C}(\rho) = S\left(\rho^A\right)-\mathop{\sup}\limits_{\left\{B_k \right\}}S\left(\rho\left|{\left\{B_k \right\}}\right.\right)
\end{align}
where $\left\{B_k \right\} (k=0,1)$ stands for the von Neumann measurements on
subsystem $B$. From the above definitions, we can see that quantum discord is
the quantumness coming from quantum measurement, while entanglement is from
the wave function (more generally density matrix) itself.


\noindent\textbf{Acknowledgements}\\
We are grateful to M. S. Sarandy for useful discussions. L.-A.W. acknowledges
for financial support from the Basque Government (Grant No. IT472-10), the
Spanish MICINN (Project No. FIS2012-36673-C03-03), and the Basque Country
University UFI (Project No. 11/55-01-2013).

\noindent\textbf{Author contributions}\\
G.-M. Z. contributed to the theoretical derivations, numerical calculations and physical analysis, L.-A. W. to the conception and design of this work. Both authors participated in the discussions and the writing of all versions of the manuscript. H.-J. X. contributed to the initial derivation, calculation and analysis and participated in the first version of the manuscript.

\noindent\textbf{Additional Information} \\
Competing financial interests: The authors declare no competing financial interests.

\noindent\textbf{Corresponding Authors}\\
Correspondence and requests for materials should be addressed to G.-M. Z. (gmzeng@jlu.edu.cn) or
L.-A. W. (lianao.wu@ehu.es).

\clearpage

\begin{figure}[htbp]
\centering
\includegraphics[width=9.0cm]{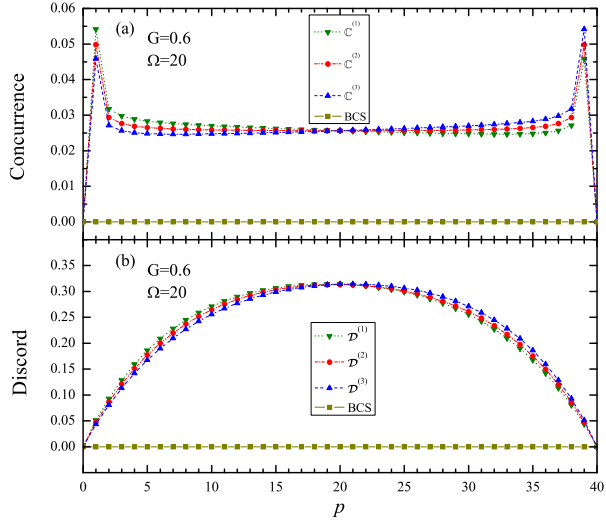}
\caption{(Color online) PBCS calculations for (a) Concurrence and (b) Quantum
discord versus particle-pair number $p$ at the paring strength $G=0.6$, single
particle energies $\epsilon_1 = 0$ and $\epsilon_2 = 1$. Both levels have the
same degeneracy $\Omega=20$. The results of BCS are also drawn as a reference.
All connection lines between points are just for guiding eyes and the same below.
\label{fig1}}
\end{figure}

\begin{figure}[htbp]
\centering
\includegraphics[width=9.0cm]{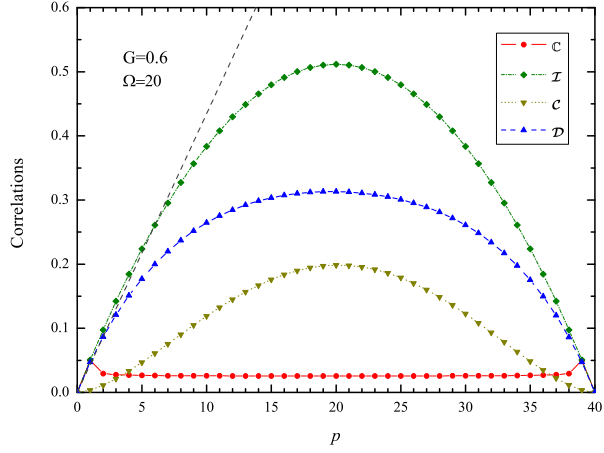}
\caption{(Color online) Discord, concurrence, total correlation, and
classical correlation versus pair number $p$ at the paring
strength $G=0.6$, and $\Omega= 20$. Single particle energies are
$\epsilon_1 = 0$ and $\epsilon_2 = 1$.} \label{fig2}
\end{figure}

\begin{figure}[htbp]
\centering
\includegraphics[width=9.0cm]{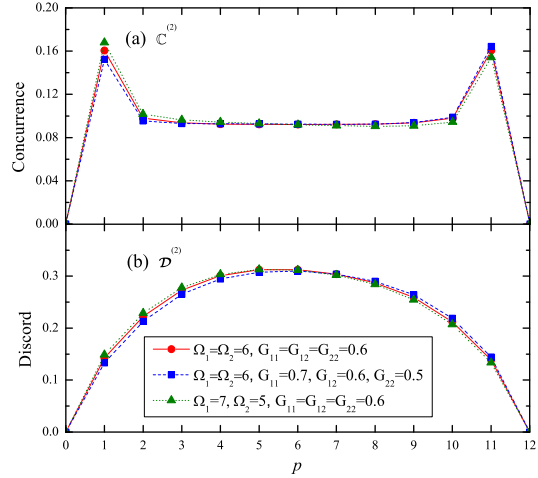}
\caption{(Color online) Concurrence and discord versus particle-pair number
$p$ from the exact solutions of Hamiltonian (\ref{eq:2}). Single particle
energies are $\epsilon_1 = 0$ and $\epsilon_2 = 1$. \label{fig3}}
\end{figure}

\end{document}